\documentclass[a4paper]{jpconf}
\usepackage{amssymb,amsmath,amsfonts}
\usepackage{graphicx}
\usepackage{graphics}
\usepackage{eepic,epsfig}
\begin{document}

\title{Fermionic Casimir effect in de Sitter spacetime}

\author{A A Saharian}

\address{Department of Physics, Yerevan State University, 1 Alex Manoogian Street, 0025 Yerevan, Armenia}

\ead{saharian@ictp.it}

\begin{abstract}
The Casimir densities are investigated for a massive spinor field
in de Sitter spacetime with an arbitrary number of toroidally
compactified spatial dimensions. The vacuum expectation value of
the energy-momentum tensor is presented in the form of the sum of
corresponding quantity in the uncompactified de Sitter spacetime
and the part induced by the non-trivial topology. The latter is
finite and the renormalization is needed for the first part only.
The asymptotic behavior of the topological term is investigated in
the early and late stages of the cosmological expansion. When the
comoving lengths of the compactified dimensions are much smaller
than the de Sitter curvature radius, to the leading order the
topological part coincides with the corresponding quantity for a
massless fermionic field and is conformally related to the
corresponding flat spacetime result with the same topology. In
this limit the topological term dominates the uncompactified de
Sitter part and the back-reaction effects should be taken into
account. In the opposite limit, for a massive field the asymptotic
behavior of the topological part is damping oscillatory.\\

\bigskip

\end{abstract}

\section{Introduction}

Many of high energy theories of fundamental physics are formulated
in higher dimensional spacetimes and it is commonly assumed that
the extra dimensions are compactified. In particular, the idea of
compactified dimensions has been extensively used in supergravity
and superstring theories. From an inflationary point of view
universes with compact spatial dimensions, under certain
conditions, should be considered a rule rather than an exception
\cite{Lind04}. These models may play an important role by
providing proper initial conditions for inflation (for physical
motivations of considering compact universes see also
\cite{Star98}). As it was argued in \cite{McIn04}, there are many
reasons to expect that in string theory the most natural topology
for the universe is that of a flat compact three-manifold. The
quantum creation of the universe having toroidal spatial topology
is discussed in \cite{Zeld84} and in references \cite{Gonc85}
within the framework of various supergravity theories. In addition
to the theoretical work, there has been a large activity to search
for signatures of non-trivial topology by identifying ghost images
of galaxies, clusters or quasars. Recent progress in observations
of the cosmic microwave background provides an alternative way to
observe the topology of the universe \cite{Levi02}. If the scale
of periodicity is close to the particle horizon scale then the
changed appearance of the microwave background sky pattern offers
a sensitive probe of the topology.

The compactification of spatial dimensions leads to a number of
interesting quantum effects which include instabilities in
interacting field theories \cite{Ford80a}, topological mass
generation \cite{Ford79,Toms80a,Toms80b} and symmetry breaking
\cite{Toms80b,Odin88}. In the case of non-trivial topology the
boundary conditions imposed on fields give rise to the
modification of the spectrum for vacuum fluctuations and, as a
result, to the Casimir-type contributions in the vacuum
expectation values of physical observables (for the topological
Casimir effect and its role in cosmology see
\cite{Most97,Bord01}). A characteristic feature of models with
compactified dimensions is the presence of moduli fields which
parametrize the size and the shape of the extra dimensions and the
Casimir effect has been used for the generation of the effective
potential for these moduli. The Casimir energy can also serve as a
model for dark energy needed for the
explanation of the present accelerated expansion of the universe (see \cite%
{Milt03,Eliz06,Gree07} and references therein).

Quantum field theory in de Sitter (dS) spacetime has been extensively
studied during the past two decades. Much of early interest was motivated by
the questions related to the quantization of fields on curved backgrounds.
dS spacetime has a high degree of symmetry and numerous physical problems
are exactly solvable on this background. The importance of this theoretical
work increased by the appearance of \ the inflationary cosmology scenario
\cite{Lind90}. In most inflationary models an approximately dS spacetime is
employed to solve a number of problems in standard cosmology. During an
inflationary epoch quantum fluctuations in the inflaton field introduce
inhomogeneities and may affect the transition toward the true vacuum. These
fluctuations play a central role in the generation of cosmic structures from
inflation. More recently astronomical observations of high redshift
supernovae, galaxy clusters and cosmic microwave background \cite{Ries07}
indicate that at the present epoch the Universe is accelerating and can be
well approximated by a world with a positive cosmological constant. If the
Universe would accelerate indefinitely, the standard cosmology would lead to
an asymptotic dS universe. Hence, the investigation of physical effects in
dS spacetime is important for understanding both the early Universe and its
future.

One-loop quantum effects for various spin fields on the background of dS
spacetime have been discussed by several authors (see, for instance, \cite%
{Cher68}-\cite{Birr82} and references therein). The effects of the
toroidal compactification of spatial dimensions in dS spacetime on
the properties of quantum vacuum for a scalar field with general
curvature coupling parameter are investigated in
\cite{Saha07,Bell08,Saha08b} (the quantum effects in braneworld
models with dS spaces and in higher-dimensional brane models with
compact internal spaces were discussed in \cite{dSbrane,Flac03}).
In the present talk, based on \cite{Saha08,Beze08} we present the
results on the Casimir effect for a fermionic field on background
of dS spacetime with spatial topology $\mathrm{R}^{p}\times
(\mathrm{S}^{1})^{q}$.

The paper is organized as follows. In the next section the plane
wave eigenspinors in $(D+1)$-dimensional dS spacetime with an
arbitrary number of toroidally compactified dimensions are
constructed. In section \ref{sec:EMT} these eigenspinors are used
for the evaluation of the vacuum expectation value of the
energy-momentum tensor in both cases of the fields with
periodicity and antiperiodicity conditions along compactified
dimensions. The behavior of these quantities is investigated in
asymptotic regions of
the parameters. The special case of topology $\mathrm{R}^{D-1}\times \mathrm{%
S}^{1}$ with the corresponding numerical results is discussed in section \ref%
{sec:Special}. The main results are summarized in section \ref{sec:Conc}.

\section{Plane wave eigenspinors in de Sitter spacetime with compactified
dimensions}

\label{sec:EigFunc}

We consider a quantum fermionic field $\psi $ on background of $(D+1)$%
-dimensional de Sitter spacetime, $\mathrm{dS}_{D+1}$, described by the line
element%
\begin{equation}
ds^{2}=dt^{2}-e^{2t/\alpha }\sum_{l=1}^{D}(dz^{l})^{2},  \label{ds2deSit}
\end{equation}%
where the parameter $\alpha $ in the expression for the scale factor is
related to the corresponding cosmological constant $\Lambda $ by the formula
$\alpha ^{2}=D(D-1)/(2\Lambda )$. We assume that the spatial coordinates $%
z^{l}$, $l=p+1,\ldots ,D$, are compactified to $\mathrm{S}^{1}$ of the
length $L_{l}$: $0\leqslant z^{l}\leqslant L_{l}$, and for the other
coordinates we have $-\infty <z^{l}<+\infty $, $l=1,\ldots ,p$. Hence, we
consider the spatial topology $\mathrm{R}^{p}\times (\mathrm{S}^{1})^{q}$,
where $q=D-p$.

The dynamics of the field is governed by the covariant Dirac equation
\begin{equation}
i\gamma ^{\mu }\nabla _{\mu }\psi -m\psi =0\ ,\;\nabla _{\mu }=\partial
_{\mu }+\Gamma _{\mu },  \label{Direq}
\end{equation}%
where $\gamma ^{\mu }=e_{(a)}^{\mu }\gamma ^{(a)}$ are the
generalized Dirac matrices and $\Gamma _{\mu }$ is the spin
connection. The latter is given in terms of the flat-space Dirac
matrices $\gamma ^{(a)}$ by
\begin{equation}
\Gamma _{\mu }=\frac{1}{4}\gamma ^{(a)}\gamma ^{(b)}e_{(a)}^{\nu }e_{(b)\nu
;\mu }\ .  \label{Gammamu}
\end{equation}%
In these relations $e_{(a)}^{\mu }$ are the tetrad components defined by $%
e_{(a)}^{\mu }e_{(b)}^{\nu }\eta ^{ab}=g^{\mu \nu }$, with $\eta ^{ab}$
being the Minkowski spacetime metric tensor. In the $(D+1)$-dimensional flat
spacetime the Dirac matrices are $N\times N$ matrices with $N=2^{[(D+1)/2]}$%
, where the square brackets mean the integer part of the enclosed
expression. We will assume that these matrices are given in the chiral
representation:
\begin{equation}
\gamma ^{(0)}=\left(
\begin{array}{cc}
1 & 0 \\
0 & -1%
\end{array}%
\right) ,\;\gamma ^{(a)}=\left(
\begin{array}{cc}
0 & \sigma _{a} \\
-\sigma _{a}^{+} & 0%
\end{array}%
\right) ,\;a=1,2,\ldots ,D.  \label{gam0l}
\end{equation}%
From the anticommutation relations for the Dirac matrices one has
$\sigma _{a}\sigma _{b}^{+}+\sigma _{b}\sigma _{a}^{+}=2\delta
_{ab} $.

In the discussion below we will denote the position vectors along the
uncompactified and compactified dimensions by $\mathbf{z}_{p}=(z^{1},\ldots
,z^{p})$ and $\mathbf{z}_{q}=(z^{p+1},\ldots ,z^{D})$. One of the
characteristic features of the field theory on backgrounds with non-trivial
topology is the appearance of inequivalent types of fields with the same
spin \cite{Isha78}. For fermion fields the boundary conditions along the
compactified dimensions can be either periodic (untwisted field) or
antiperiodic (twisted field). First we consider the field with periodicity
conditions (no summation over $l$):%
\begin{equation}
\psi (t,\mathbf{z}_{p},\mathbf{z}_{q}+L_{l}\mathbf{e}_{l})=\psi (t,\mathbf{z}%
_{p},\mathbf{z}_{q}),  \label{bc}
\end{equation}%
where $l=p+1,\ldots ,D$ and $\mathbf{e}_{l}$ is the unit vector
along the direction of the coordinate $z^{l}$. The case of a
fermionic field with antiperiodicity conditions will be discussed
below. We are interested in the effects of non-trivial topology on
the vacuum expectation value (VEV) of the energy-momentum tensor
of the fermionic field assuming that the field is prepared in the
Bunch-Davies vacuum state (also called Euclidean vacuum). In order
to evaluate this VEV, we will use the direct mode-summation
procedure. In this approach the knowledge of the complete set of
properly normalized eigenspinors, $\{\psi _{\beta }^{(+)},\psi
_{\beta }^{(-)}\}$, is needed. Here the collective index $\beta $
presents a set of quantum numbers specifying the solutions. By
virtue of spatial translation invariance the spatial part of the
eigenfunctions $\psi _{\beta }^{(\pm )}$ can be taken in the
standard plane wave form $e^{\pm i\mathbf{k z}}$. We will
decompose the wave vector into the components along the
uncompactified and compactified dimensions, $\mathbf{k=}(\mathbf{k}_{p},%
\mathbf{k}_{q})$ with
\begin{equation}
k=\sqrt{\mathbf{k}_{p}^{2}+\mathbf{k}_{q}^{2}}.  \label{ka}
\end{equation}%
For a spinor field with periodicity conditions along the compactified
dimensions the corresponding wave vector has the components%
\begin{equation}
\mathbf{k}_{q}=(2\pi n_{p+1}/L_{p+1},\ldots ,2\pi n_{D}/L_{D}),  \label{kq}
\end{equation}%
where $n_{p+1},\ldots ,n_{D}=0,\pm 1,\pm 2,\ldots $.

Choosing the basis tetrad in the form%
\begin{equation}
e_{\mu }^{(0)}=\delta _{\mu }^{0},\;e_{\mu }^{(a)}=e^{t/\alpha }\delta _{\mu
}^{a},\;a=1,2,\ldots ,D,  \label{Tetrad}
\end{equation}
for the positive frequency plane wave eigenspinors corresponding
to the Bunch-Davies vacuum and satisfying the periodicity
conditions along the compactified dimensions we have
\begin{equation}
\psi _{\beta }^{(+)}=A_{\beta }\eta ^{(D+1)/2}e^{i\mathbf{k}\cdot \mathbf{r}%
}\left(
\begin{array}{c}
H_{1/2-i\alpha m}^{(1)}(k\eta )w_{\sigma }^{(+)} \\
-i(\mathbf{n}\cdot \boldsymbol{\sigma })H_{-1/2-i\alpha m}^{(1)}(k\eta
)w_{\sigma }^{(+)}%
\end{array}%
\right) ,  \label{psibet+n}
\end{equation}%
where $\beta =(\mathbf{k},\sigma )$, $\mathbf{n}=\mathbf{k}/k$, and $%
w_{\sigma }^{(+)}$, $\sigma =1,\ldots ,N/2$, are one-column matrices having $%
N/2$ rows with the elements $w_{l}^{(\sigma )}=\delta _{l\sigma }$. In (\ref%
{psibet+n}), $H_{\nu }^{(1)}(x)$ is the Hankel function, $%
\boldsymbol{\sigma
}=(\sigma _{1},\sigma _{2},\ldots ,\sigma _{D})$, and we are using the
notation%
\begin{equation}
\eta =\alpha e^{-t/\alpha },\;0\leqslant \eta <\infty .  \label{etavar}
\end{equation}%
Note that $\tau =-\eta $ is the conformal time coordinate, in terms of which
the dS line element (\ref{ds2deSit}) takes the conformally flat form.
Similarly, the negative frequency eigenspinors have the form
\begin{equation}
\psi _{\beta }^{(-)}=A_{\beta }\eta ^{(D+1)/2}e^{-i\mathbf{k}\cdot \mathbf{r}%
}\left(
\begin{array}{c}
i(\mathbf{n}\cdot \boldsymbol{\sigma })H_{-1/2+i\alpha m}^{(2)}(k\eta
)w_{\sigma }^{(-)} \\
H_{1/2+i\alpha m}^{(2)}(k\eta )w_{\sigma }^{(-)}%
\end{array}%
\right) ,  \label{psibet-}
\end{equation}%
with $w_{\sigma }^{(-)}=iw_{\sigma }^{(+)}$.

The coefficients $A_{\beta }$ in the expressions for the eigenspinors are
determined from the orthonormalization condition%
\begin{equation}
\int d^{D}x\,\sqrt{\gamma }\psi _{\beta }^{(\pm )+}\psi _{\beta ^{\prime
}}^{(\pm )}=\delta _{\beta \beta ^{\prime }},  \label{normaliz}
\end{equation}%
with $\gamma $ being the determinant of the spatial metric. On the
right-hand side of this condition the symbol $\delta _{\beta \beta ^{\prime
}}$ is understood as the Dirac delta function for the continuous indices and
the Kronecker delta for the discrete ones. By making use of the Wronskian
relation for the Hankel functions, we can see that%
\begin{equation}
A_{\beta }^{2}=\frac{ke^{\pi \alpha m}}{2^{p+2}\pi ^{p-1}V_{q}\alpha ^{D}},
\label{Abet}
\end{equation}%
where $V_{q}=L_{p+1}\cdots L_{D}$ is the volume of the compactified
subspace. For a massless fermionic field the Hankel functions in (\ref%
{psibet+n}) and (\ref{psibet-}) are expressed in terms of exponentials and
we have the standard conformal relation $\psi _{\beta }^{(\pm )}=(\eta
/\alpha )^{(D+1)/2}\psi _{\beta }^{\mathrm{(M)}(\pm )}$ between eigenspinors
defining the Bunch-Davies vacuum in dS spacetime and the corresponding
eigenspinors $\psi _{\beta }^{\mathrm{(M)}(\pm )}$ for the Minkowski
spacetime with spatial topology $\mathrm{R}^{p}\times (\mathrm{S}^{1})^{q}$.

The plane wave eigenspinors for a twisted spinor field are constructed in a
similar way. For this field we have antiperiodicity conditions along the
compactified dimensions:%
\begin{equation}
\psi (t,\mathbf{z}_{p},\mathbf{z}_{q}+L_{l}\mathbf{e}_{l})=-\psi (t,\mathbf{z%
}_{p},\mathbf{z}_{q}).  \label{AntCond}
\end{equation}%
The corresponding eigenspinors are given by expressions (\ref{psibet+n}), (%
\ref{psibet-}), where now the components of the wave vector along the
compactified dimensions are given by the formula
\begin{equation}
\mathbf{k}_{q}=(\pi (2n_{p+1}+1)/L_{p+1},\ldots ,\pi (2n_{D}+1)/L_{D}),
\label{kqCompTw}
\end{equation}%
with $n_{p+1},\ldots ,n_{D}=0,\pm 1,\pm 2,\ldots $, and
\begin{equation}
k^{2}=\mathbf{k}_{p}^{2}+\sum_{l=p+1}^{D}\left[ \pi (2n_{l}+1)/L_{l}\right]
^{2}.  \label{kTw}
\end{equation}%
Note that the physical wave vector is given by the combination $\mathbf{k}%
\eta /\alpha $.

\section{Vacuum expectation value of the energy-momentum tensor}

\label{sec:EMT}

The compactification of the spatial dimensions leads to the modification of
the spectrum for zero-point fluctuations of fields and as a result of this
the VEVs of physical observables are changed. Among the most important
quantities characterizing the properties of the vacuum is the expectation
value of the energy-momentum tensor. In addition to describing the physical
structure of the quantum field at a given point, the energy-momentum tensor
acts as a source of gravity in the Einstein equations and plays an important
role in modelling a self-consistent dynamics involving the gravitational
field. In order to study the one-loop topological effects in the VEV of the
energy-momentum tensor we will use the direct mode summation approach. The
corresponding mode-sum has the form%
\begin{equation}
\langle 0|T_{\mu \nu }|0\rangle =\frac{i}{2}\int d\mathbf{k}_{p}\sum_{%
\mathbf{k}_{q},\sigma }[\bar{\psi}_{\mathbf{k},\sigma }^{(-)}\gamma _{(\mu
}\nabla _{\nu )}\psi _{\mathbf{k},\sigma }^{(-)}-(\nabla _{(\mu }\bar{\psi}_{%
\mathbf{k},\sigma }^{(-)})\gamma _{\nu )}\psi _{\mathbf{k},\sigma }^{(-)}],
\label{VEVEMT}
\end{equation}%
where the eigenspinors $\psi _{\beta }^{(-)}=\psi _{\mathbf{k},\sigma }^{(-)}
$ are given by (\ref{psibet-}) and $\bar{\psi}_{\mathbf{k},\sigma
}^{(-)}=\psi _{\mathbf{k},\sigma }^{(-)+}\gamma ^{0}$ is the Dirac adjoint.
For a fermionic field with periodic boundary conditions, substituting the
eigenspinors (\ref{psibet-}) into this formula, for the energy density and
the vacuum stresses we find (no summation over $l$)%
\begin{eqnarray}
\langle 0|T_{0}^{0}|0\rangle  &=&\frac{2^{-p}\eta ^{D+1}\alpha ^{-D-1}}{\pi
^{p/2-1}\Gamma (p/2)V_{q}}\int_{0}^{\infty
}dk_{p}\,k_{p}^{p-1}\sum_{n_{p+1}=-\infty }^{+\infty }\cdots
\sum_{n_{D}=-\infty }^{+\infty }k^{2}  \notag \\
&&\times {\mathrm{Re}}\left[ H_{1/2+i\alpha m}^{(1)}(k\eta )H_{-1/2+i\alpha
m}^{(2)\prime }(k\eta )-H_{-1/2+i\alpha m}^{(1)}(k\eta )H_{1/2+i\alpha
m}^{(2)\prime }(k\eta )\right] ,  \label{T00} \\
\langle 0|T_{l}^{l}|0\rangle  &=&\frac{2^{1-p}\eta ^{D+1}\alpha ^{-D-1}}{\pi
^{p/2-1}\Gamma (p/2)V_{q}}\int_{0}^{\infty
}dk_{p}\,k_{p}^{p-1}\sum_{n_{p+1}=-\infty }^{+\infty }\cdots
\sum_{n_{D}=-\infty }^{+\infty }k_{l}^{2}  \notag \\
&&\times {\mathrm{Re}}\left[ H_{1/2+i\alpha m}^{(1)}(k\eta )H_{1/2+i\alpha
m}^{(2)}(k\eta )\right] ,  \label{Tll}
\end{eqnarray}%
with $l=1,2,\ldots ,D$, and the off-diagonal components vanish. The VEVs
given by (\ref{T00}) and (\ref{Tll}) are divergent and need some
regularization procedure. To make them finite we can introduce a cut-off
function $\varphi _{\lambda }(k)$ with the cut-off parameter $\lambda $,
which decreases sufficiently fast with increasing $k$ and satisfies the
condition $\varphi _{\lambda }(k)\rightarrow 1$, for $\lambda \rightarrow 0$.

As the next step, we apply to the series over $n_{p+1}$ in (\ref{T00}) and (%
\ref{Tll}) the Abel-Plana summation formula \cite{Most97,Saha07Gen}%
\begin{equation}
\sideset{}{'}{\sum}_{n=0}^{\infty }f(n)=\int_{0}^{\infty
}dx\,f(x)+i\int_{0}^{\infty }dx\,\frac{f(ix)-f(-ix)}{e^{2\pi x}-1},
\label{Abel}
\end{equation}%
where the prime means that the term $n=0$ should be halved. The term in the
VEV with the first integral in the right-hand side of this formula
corresponds to the energy-momentum tensor in dS spacetime with topology $%
\mathrm{R}^{p+1}\times (\mathrm{S}^{1})^{q-1}$. As a result, the VEV of the
energy-momentum tensor is presented in the decomposed form%
\begin{equation}
\langle T_{k}^{l}\rangle _{p,q}=\langle T_{k}^{l}\rangle _{p+1,q-1}+\Delta
_{p+1}\langle T_{k}^{l}\rangle _{p,q}.  \label{TllDecomp}
\end{equation}%
Here $\langle T_{k}^{l}\rangle _{p+1,q-1}$ is the VEV of the energy-momentum
tensor for the topology $\mathrm{R}^{p+1}\times (\mathrm{S}^{1})^{q-1}$ and
the part (no summation over $l$)%
\begin{eqnarray}
\Delta _{p+1}\langle T_{l}^{l}\rangle _{p,q} &=&\frac{N\eta ^{D+2}L_{p+1}}{%
(2\pi )^{(p+1)/2}V_{q}\alpha ^{D+1}}\sum_{n=1}^{\infty
}\sum_{n_{p+2}=-\infty }^{+\infty }\cdots \sum_{n_{D}=-\infty }^{+\infty
}\int_{0}^{\infty }dx\,x  \notag \\
&&\times \frac{{\mathrm{Re}}[I_{-1/2-i\alpha m}^{2}(\eta x)-I_{1/2+i\alpha
m}^{2}(\eta x)]}{\cosh (\alpha m\pi )(L_{p+1}n)^{p+1}}f_{p}^{(l)}(nL_{p+1}%
\sqrt{x^{2}+k_{\mathbf{n}_{q-1}}^{2}}),  \label{TopTll}
\end{eqnarray}%
is due to the compactness of the $(p+1)$th dimension. In (\ref{TopTll}), $%
I_{\nu }(x)$ and $K_{\nu }(x)$ are the modified Bessel functions and we have
introduced the notations%
\begin{equation}
f_{\nu }(x)\equiv x^{\nu }K_{\nu }(x),\;k_{\mathbf{n}_{q-1}}^{2}=%
\sum_{l=p+2}^{D}(2\pi n_{l}/L_{l})^{2}.  \label{kndmin1}
\end{equation}%
For the separate components the functions $f_{p}^{(l)}(y)$ in
formula (\ref{TopTll}) have the form
\begin{eqnarray}
f_{p}^{(l)}(y) &=&f_{(p+1)/2}(y),\;l=0,1,\ldots ,p,  \notag \\
f_{p}^{(p+1)}(y) &=& -pf_{(p+1)/2}(y)-y^{2}f_{(p-1)/2}(y),
\label{fpl} \\
f_{p}^{(l)}(y) &=&k_{l}^{2}(nL_{p+1})^{2}f_{(p-1)/2}(y),\;l=p+2,\ldots ,D,
\notag
\end{eqnarray}%
where $k_{l}=2\pi n_{l}/L_{l}$. From (\ref{TopTll}) we see that
the topological part depends on the variable $\eta $ and the
length scales $L_{l} $ in the combinations $L_{l}/\eta $. Noting
that $a(\eta )L_{l}$ is the comoving length with $a(\eta )=\alpha
/\eta $ being the scale factor, we conclude that the topological
part of the energy-momentum tensor is a function of comoving
lengths of the compactified dimensions.

The topological term (\ref{TopTll}) is finite and by the recurrence relation
(\ref{TllDecomp}) the renormalization procedure for the VEV\ of the
energy-momentum tensor is reduced to the renormalization of the
corresponding VEV in uncompactified dS spacetime. It can be seen that the
topological part is traceless for a massless field and is covariantly
conserved: $\left( \Delta _{p+1}\langle T_{l}^{k}\rangle _{p,q}\right)
_{;k}=0$. From the first relation in (\ref{fpl}) we see that the vacuum
stresses along the uncompactified dimensions are equal to the energy density,%
\begin{equation}
\Delta _{p+1}\langle T_{0}^{0}\rangle _{p,q}=\Delta _{p+1}\langle
T_{1}^{1}\rangle _{p,q}=\cdots =\Delta _{p+1}\langle T_{p}^{p}\rangle _{p,q},
\label{T00T11}
\end{equation}%
and, hence, in the uncompactified subspace the equation of state for the
topological part of the energy-momentum tensor is of the cosmological
constant type. Note that the topological parts are time-dependent and they
break the dS symmetry.

For a massless fermionic field, the modified Bessel functions in (\ref%
{TopTll}) are expressed in terms of elementary functions and after the
integration over $x$ one finds (no summation over $l$)%
\begin{equation}
\Delta _{p+1}\langle T_{l}^{l}\rangle _{p,q}=\frac{2N(\eta /\alpha )^{D+1}}{%
(2\pi )^{p/2+1}V_{q}L_{p+1}^{p+1}}\sum_{n=1}^{\infty }\sum_{n_{p+2}=-\infty
}^{+\infty }\cdots \sum_{n_{D}=-\infty }^{+\infty }\frac{%
g_{p}^{(l)}(nL_{p+1}k_{\mathbf{n}_{q-1}})}{n^{p+2}},  \label{DelTConf}
\end{equation}%
with the notations%
\begin{eqnarray}
g_{p}^{(l)}(y) &=&f_{p/2+1}(y),\;l=0,1,\ldots ,p,  \notag \\
g_{p}^{(p+1)}(y) &=&-(p+1)f_{p/2+1}(y)-y^{2}f_{p/2}(y),  \label{gi} \\
g_{p}^{(l)}(y) &=&(nL_{p+1}k_{l})^{2}f_{p/2}(y),\;l=p+2,\ldots ,D.  \notag
\end{eqnarray}%
The massless fermionic field is conformally invariant in any dimension and
in this case the problem under consideration is conformally related to the
corresponding problem in the Minkowski spacetime with spatial topology $%
\mathrm{R}^{p}\times (\mathrm{S}^{1})^{q}$. Formula
(\ref{DelTConf}) could also be obtained from the relation $\Delta
_{p+1}\langle T_{i}^{i}\rangle
_{p,q}=a^{-D-1}(\eta )\Delta _{p+1}\langle T_{i}^{i}\rangle _{p,q}^{\mathrm{%
(M)}}$, with $a(\eta )$ being the scale factor. Comparing expression (\ref%
{DelTConf}) with the corresponding formula from \cite{Bell08} for a
conformally coupled massless scalar field, we see that the following
relation takes place: $\Delta _{p+1}\langle T_{k}^{l}\rangle _{p,q}=-N\Delta
_{p+1}\langle T_{k}^{l}\rangle _{p,q}^{\mathrm{(scalar)}}$.

After the repetitive application of the recurrence formula (\ref{TllDecomp}%
), the VEV of the energy-momentum tensor for the topology $\mathrm{R}%
^{p}\times (\mathrm{S}^{1})^{q}$ is presented in the form%
\begin{equation}
\langle T_{l}^{k}\rangle _{p,q}=\langle T_{l}^{k}\rangle _{\mathrm{dS,ren}%
}+\langle T_{l}^{k}\rangle _{\mathrm{c}},\;\langle T_{l}^{k}\rangle _{%
\mathrm{c}}=\sum_{l=1}^{q}\Delta _{D+1-l}\langle T_{l}^{k}\rangle _{D-l,l},
\label{TlkdS}
\end{equation}%
where $\langle T_{l}^{k}\rangle _{\mathrm{dS,ren}}$ is the renormalized VEV
in the uncompactified dS spacetime and $\langle T_{l}^{k}\rangle _{\mathrm{c}%
}$ is the topological part. For $D=3$ the renormalized VEV of the
energy-momentum tensor for fermionic field in uncompactified dS spacetime is
investigated in \cite{Mama81} (see also \cite{Most97}). The corresponding
expression has the form%
\begin{equation}
\langle T_{l}^{k}\rangle _{\mathrm{dS,ren}}^{(D=3)}=\frac{\delta _{l}^{k}}{%
16\pi ^{2}\alpha ^{4}}\left\{ 2m^{2}\alpha ^{2}(m^{2}\alpha ^{2}+1)[\ln
(m\alpha )-{\mathrm{Re\,}}\psi (im\alpha )]+m^{2}\alpha ^{2}/6+11/60\right\}
,  \label{TkldSren3}
\end{equation}%
where $\psi (x)$ is the logarithmic derivative of the gamma-function. In the
limit of zero mass, only the last term in curly brackets contributes and the
corresponding energy density is positive. For large values of the mass we
have:%
\begin{equation}
\langle T_{l}^{k}\rangle _{\mathrm{dS,ren}}^{(D=3)}\approx -\frac{\delta
_{l}^{k}}{960\pi ^{2}\alpha ^{6}m^{2}},\;m\alpha \gg 1,  \label{largeMassD3}
\end{equation}%
and the energy density is negative.

\begin{figure}[tbph]
\begin{center}
\epsfig{figure=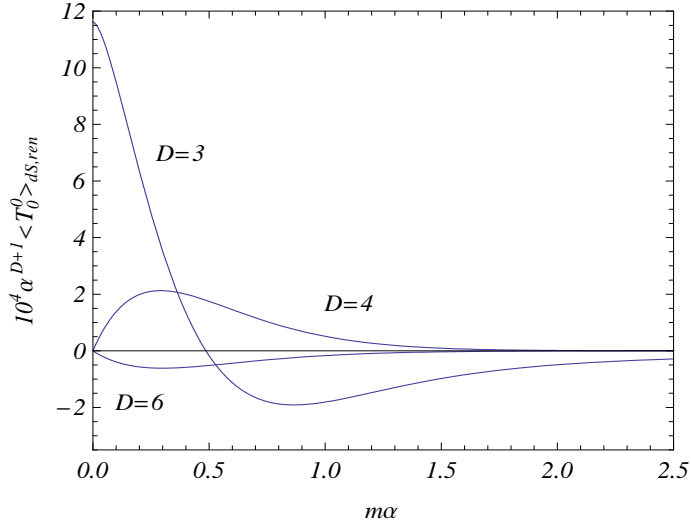,width=9.cm,height=7.cm}
\end{center}
\caption{The vacuum energy density in uncompactified dS spacetime, $\langle
T_{0}^{0}\rangle _{\mathrm{dS,ren}}$, as a function of the parameter $m%
\protect\alpha $ for spatial dimensions $D=3,4,6$. }
\label{fig1}
\end{figure}

In higher dimensions a closed expression can be obtained for even values of $D$%
. In this case for the renormalized VEV of the energy-momentum tensor in
uncompactified dS spacetime we have the formula \cite{Beze08}%
\begin{equation}
\langle T_{l}^{k}\rangle _{\mathrm{dS,ren}}=\frac{2Nm^{D+1}\delta _{l}^{k}}{%
(4\pi )^{(D+1)/2}(D+1)}\frac{\Gamma \left( (1-D)/2\right) }{e^{2\pi m\alpha
}+1}\prod\limits_{j=0}^{D/2-1}\bigg[1+\frac{(j+1/2)^{2}}{m^{2}\alpha ^{2}}%
\bigg].  \label{TkldSren}
\end{equation}%
For a massless fermionic field this tensor vanishes. We could expect this
result, since in odd-dimensional spacetimes the trace anomaly is absent. For
large values of the mass, $m\alpha \gg 1$, the VEV for the energy-momentum
tensor is exponentially suppressed. This is in contrast with the case of
even dimensional spacetimes, where the suppression in the large mass limit
is power-law. In particular, for 5-dimensional de Sitter spacetime formula (%
\ref{TkldSren}) leads to the result%
\begin{equation}
\langle T_{l}^{k}\rangle _{\mathrm{dS,ren}}=\frac{m^{5}\delta _{l}^{k}}{%
15\pi ^{2}}\frac{1}{e^{2\pi m\alpha }+1}\left( 1+\frac{5}{2m^{2}\alpha ^{2}}+%
\frac{9}{16m^{4}\alpha ^{4}}\right) ,\;D=4.  \label{TDS5}
\end{equation}%
The corresponding energy density is positive. In figure \ref{fig1}
we have plotted the vacuum energy density $\langle
T_{0}^{0}\rangle _{\mathrm{dS,ren}}$ as a function of the
parameter $m\alpha $ for spatial dimensions $D=3,4,6$.

In uncompactified dS spacetime the energy-momentum tensor $\langle
T_{l}^{k}\rangle _{\mathrm{dS,ren}}$ is time-independent and corresponds to
a gravitational source of the cosmological constant type. These properties
directly follow from the dS invarinace of the Bunch-Davies vacuum state.
Combining with the initial cosmological constant $\Lambda $, one-loop
effects in uncompactified dS spacetime lead to the effective cosmological
constant
\begin{equation}
\Lambda _{\mathrm{eff}}=\Lambda +8\pi G_{D+1}\langle T_{0}^{0}\rangle _{%
\mathrm{dS,ren}},  \label{effCC}
\end{equation}%
where $G_{D+1}$ is the gravitational constant in $(D+1)$-dimensional
spacetime. In the discussion above we have assumed that the quantum state of
a fermionic field is the Bunch-Davies vacuum state. In \cite{Ande00} it was
shown that for a scalar field with a wide range of mass and curvature
coupling parameter the expectation values of the energy-momentum tensor in
arbitrary physically admissable states approaches the expectation value in
the Bunch-Davies vacuum at late times.

Now we return to the consideration of the topological part in the
VEV of the energy-momentum tensor and discuss its behavior in the
asymptotic regions of the parameters. For small values of the
comoving length $a(\eta )L_{p+1}$ with respect to the dS curvature
radius, $a(\eta )L_{p+1}\ll \alpha $, to the leading order the
topological part in the VEV of the energy-momentum
tensor coincides with that for a massless field given by formula (\ref%
{DelTConf}). In particular, the topological part of the vacuum energy
density is positive. This limit corresponds to the early stages of the
cosmological evolution, $t\rightarrow -\infty $, and at these stages the
total VEV is dominated by the topological part. In this limit the
back-reaction effects of the topological terms are important and these
effects can change the dynamics essentially (for a discussion of
back-reaction effects from vacuum fluctuations on the dynamics of dS
spacetime see, for instance, \cite{Pere07} and references therein).

For large values of the comoving length, $a(\eta )L_{p+1}\gg \alpha $, the
leading term in the topological part is given by the expression (no
summation over $l$):%
\begin{equation}
\Delta _{p+1}\langle T_{l}^{l}\rangle _{p,q}\approx \frac{NB_{l}\cos
[2mt-2\alpha m\ln (\alpha /L_{p+1})-\phi _{l}]}{2^{p/2}\pi
^{(p+1)/2}V_{q}L_{p+1}^{p+1}\cosh (\alpha m\pi )e^{(D+1)t/\alpha }}.
\label{TllSmall}
\end{equation}%
Here the coefficients $B_{l}$ and the phases $\phi _{l}$ are defined by the
formula
\begin{equation}
B_{l}e^{i\phi _{l}}=\frac{2^{-i\alpha m}}{\Gamma (1/2+i\alpha m)}%
\sum_{n=1}^{\infty }\sum_{n_{p+2}=-\infty }^{+\infty }\cdots
\sum_{n_{D}=-\infty }^{+\infty }\frac{h_{p}^{(l)}(nL_{p+1}k_{\mathbf{n}%
_{q-1}})}{n^{p+2+2i\alpha m}},  \label{B0}
\end{equation}%
with the notations%
\begin{eqnarray}
h_{p}^{(l)}(x) &=&f_{p/2+1+i\alpha m}(x),\;l=0,1,\ldots ,p,  \notag \\
h_{p}^{(p+1)}(x) &=&-(p+1+2i\alpha m)f_{p/2+1+i\alpha
m}(x)-x^{2}f_{p/2+i\alpha m}(x),  \label{hpl} \\
h_{p}^{(l)}(x) &=&(L_{p+1}k_{l}n)^{2}f_{p/2+i\alpha m}(x),\;l=p+2,\ldots ,D.
\notag
\end{eqnarray}
Relation (\ref{TllSmall}) describes the asymptotic behavior of the
topological part in the late stages of cosmological evolution corresponding
to the limit $t\rightarrow +\infty $. In this limit the behavior of the
topological part for a massive spinor field is damping oscillatory. The
damping factor in the amplitude and the oscillation frequency are the same
for all terms in the sum of (\ref{TlkdS}) and the total topological term
behaves as%
\begin{equation}
\langle T_{l}^{k}\rangle _{\mathrm{c}}\propto \delta
_{l}^{k}e^{-(D+1)t/\alpha }\cos \left( 2mt+\phi _{\mathrm{c}}^{\prime
}\right) ,\;t\rightarrow +\infty .  \label{Tlktlarge}
\end{equation}
As the vacuum energy-momentum tensor for uncompactified dS spacetime is
time-independent, we have similar damping oscillations in the total
energy-momentum tensor $\langle T_{l}^{k}\rangle _{\mathrm{dS,ren}}+\langle
T_{l}^{k}\rangle _{\mathrm{c}}$. This type of oscillations are absent in the
case of a massless field when the topological parts decay monotonically as $%
e^{-(D+1)t/\alpha }$. Note that in the case of a scalar field the vanishing
of the topological part is monotonic or oscillatory in dependence of the
mass and the curvature coupling parameter of the field \cite{Saha07,Bell08}.

For a fermionic field with antiperiodicity conditions (\ref{AntCond}), the
VEV\ of the energy-momentum tensor is found in a way similar to that for the
periodicity conditions. The corresponding formulae for the topological parts
are obtained from those for the field with periodicity conditions inserting
the factor $(-1)^{n}$ in the summation over $n$ and replacing the definition
for $k_{\mathbf{n}_{q-1}}^{2}$ by
\begin{equation}
k_{\mathbf{n}_{q-1}}^{2}=\sum_{l=p+2}^{D}\left[ \pi (2n_{l}+1)/L_{l}\right]
^{2}.  \label{knq-1Tw}
\end{equation}%
In situations where the main contribution comes from the term with $n=1$,
the topological parts in the VEV of the energy-momentum tensor for fields
with periodicity and antiperiodicity conditions have opposite signs.

\section{Special case}

\label{sec:Special}

In the special case of a single compactified dimension with the length $L$
(topology $\mathrm{R}^{D-1}\times \mathrm{S}^{1}$), for the topological part
in the VEV of the energy-momentum tensor for an untwisted field we have (no
summation over $l$)%
\begin{eqnarray}
\langle T_{l}^{l}\rangle _{\mathrm{c}} &=&\frac{N(\eta /L)^{D}}{(2\pi
)^{D/2}\alpha ^{D+1}\cosh (\alpha m\pi )}\sum_{n=1}^{\infty }\frac{1}{n^{D}}%
\int_{0}^{\infty }dx\,x  \notag \\
&&\times f_{D-1}^{(l)}(nxL/\eta ){\mathrm{Re}}[I_{-1/2-i\alpha
m}^{2}(x)-I_{1/2+i\alpha m}^{2}(x)].  \label{TlkSpec}
\end{eqnarray}%
For a massless fermionic field this formula reduces to (no
summation over $l$)%
\begin{equation}
\langle T_{l}^{l}\rangle _{\mathrm{c}}=\frac{N\zeta (D+1)}{\pi ^{(D+1)/2}}%
\left( \frac{\eta }{\alpha L}\right) ^{D+1}\Gamma \left( \frac{D+1}{2}%
\right) ,\;\langle T_{D}^{D}\rangle _{\mathrm{c}}=-D\langle T_{0}^{0}\rangle
_{\mathrm{c}},  \label{T00Spm0}
\end{equation}%
with $l=0,1,\ldots ,D-1$ and $\zeta (x)$ being the Riemann zeta
function. In accordance with the asymptotic analysis given before,
the expressions (\ref{T00Spm0}) describe the asymptotic behavior
of the topological part in the VEV of the energy-momentum tensor
for a massive field at early stages of the cosmological expansion.
The asymptotics at late stages are obtained from general formula
(\ref{TllSmall}) and have the form
(no summation over $l$)%
\begin{equation}
\langle T_{l}^{l}\rangle _{\mathrm{c}}\approx NC_{l}\frac{\cos [2mt-2\alpha
m\ln (\alpha /L)-\phi _{l}]}{\pi ^{D/2}\cosh (\alpha m\pi )(Le^{t/\alpha
})^{D+1}},  \label{TllSpLarge}
\end{equation}%
where now $C_{l}$ and $\phi _{l}$ are defined by the relations%
\begin{eqnarray}
C_{l}e^{i\phi _{l}} &=&\frac{\Gamma ((D+1)/2+i\alpha m)}{\Gamma (1/2+i\alpha
m)}\zeta (D+1+2i\alpha m),\;l=0,1,\ldots ,D-1,  \notag \\
C_{D}e^{i\phi _{D}} &=&-(D+2i\alpha m)C_{0}e^{i\phi _{0}}.
\label{BDSp}
\end{eqnarray}%
For a massless field, (\ref{TllSpLarge}) is reduced to the exact result (\ref%
{T00Spm0}). The corresponding formulae for a twisted spinor field are
obtained by adding the coefficient $(2^{-D}-1)$ in the first formula of (\ref%
{T00Spm0}) and the coefficient $(2^{-D-2i\alpha m}-1)$ in the first formula
of (\ref{BDSp}).

For $D=3$ the functions $f_{D-1}^{(l)}(x)$ in (\ref{TlkSpec}) are expressed
in terms of exponentials and after the summation over $n$ we find (no
summation over $l$)%
\begin{equation}
\langle T_{l}^{l}\rangle _{\mathrm{c}}=\frac{(\eta /L)^{3}}{\pi \alpha
^{4}\cosh (\alpha m\pi )}\int_{0}^{\infty }dx\,xG_{l}(Lx/\eta ){\mathrm{Re}}%
[I_{-1/2-i\alpha m}^{2}(y)-I_{1/2+i\alpha m}^{2}(y)],  \label{TllR2S1}
\end{equation}%
where the following notations are introduced%
\begin{eqnarray}
&&G_{0}(y)=y^{2}\ln (1-e^{-y}),\;G_{3}(y)=G_{0}(y)-2G_{1}(y),  \notag \\
&&G_{1}(y)=G_{2}(y)=y\mathrm{Li}_{2}(e^{-y})+\mathrm{Li}_{3}(e^{-y}),
\label{Gl}
\end{eqnarray}%
with $\mathrm{Li}_{n}(z)=\sum_{k=1}^{\infty }z^{k}/k^{n}$ being the
polylogarithm function. In particular, for a massless fermionic field we have%
\begin{equation}
\langle T_{k}^{l}\rangle _{\mathrm{c}}=\frac{2\pi ^{2}}{45}\left( \frac{\eta
}{\alpha L}\right) ^{4}\mathrm{diag}(1,1,1,-3).  \label{TlkD3m0}
\end{equation}

In the case of a twisted field and for the topology $\mathrm{R}^{2}\times
\mathrm{S}^{1}$ one has (no summation over $l$)%
\begin{equation}
\langle T_{l}^{l}\rangle _{\mathrm{c}}=\frac{(\eta /L)^{3}}{\pi \alpha
^{4}\cosh (\alpha m\pi )}\int_{0}^{\infty }dx\,xG_{l}^{\mathrm{(tw)}%
}(Lx/\eta ){\mathrm{Re}}[I_{-1/2-i\alpha m}^{2}(y)-I_{1/2+i\alpha m}^{2}(y)],
\label{TllTwR2S1}
\end{equation}%
with the notations%
\begin{eqnarray}
&&G_{0}^{\mathrm{(tw)}}(y)=y^{2}\ln (1+e^{-y}),\;G_{3}^{\mathrm{(tw)}%
}(y)=G_{0}^{\mathrm{(tw)}}(y)-2G_{l}^{\mathrm{(tw)}}(y),  \notag \\
&&G_{l}^{\mathrm{(tw)}}(y)=y\mathrm{Li}_{2}(-e^{-y})+\mathrm{Li}%
_{3}(-e^{-y}),\;l=1,2.  \label{Gltw}
\end{eqnarray}%
For a massless twisted field one has the result%
\begin{equation}
\langle T_{k}^{l}\rangle _{\mathrm{c}}=-\frac{7\pi ^{2}}{180}\left( \frac{%
\eta }{\alpha L}\right) ^{4}\mathrm{diag}(1,1,1,-3).  \label{TlkD3m0tw}
\end{equation}%
Note that for a massless field in $\mathrm{dS}_{5}$ having spatial topology $%
\mathrm{R}^{3}\times \mathrm{S}^{1}$ the topological part in the VEV of the
energy-momentum tensor has the form%
\begin{equation}
\langle T_{k}^{l}\rangle _{\mathrm{c},m=0}=\frac{b\zeta (5)}{\pi ^{2}}\left(
\frac{\eta }{\alpha L}\right) ^{5}\mathrm{diag}(1,1,1,1,-4),
\label{TlkR2S1m0}
\end{equation}%
where $b=3$ for the field with periodicity condition and
$b=-45/16$ for the field with antiperiodicity condition. This
topology corresponds to the original Kaluza-Klein model.

\begin{figure}[tbph]
\begin{center}
\begin{tabular}{cc}
\epsfig{figure=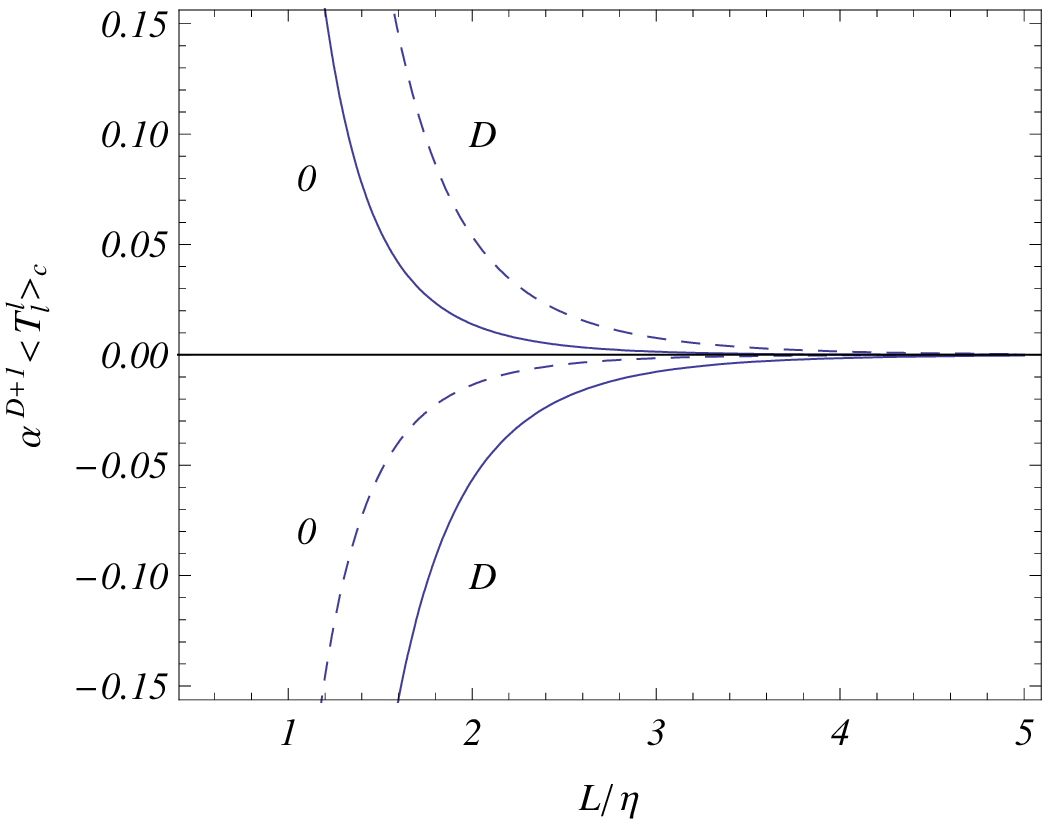,width=7.cm,height=6.cm} & \quad %
\epsfig{figure=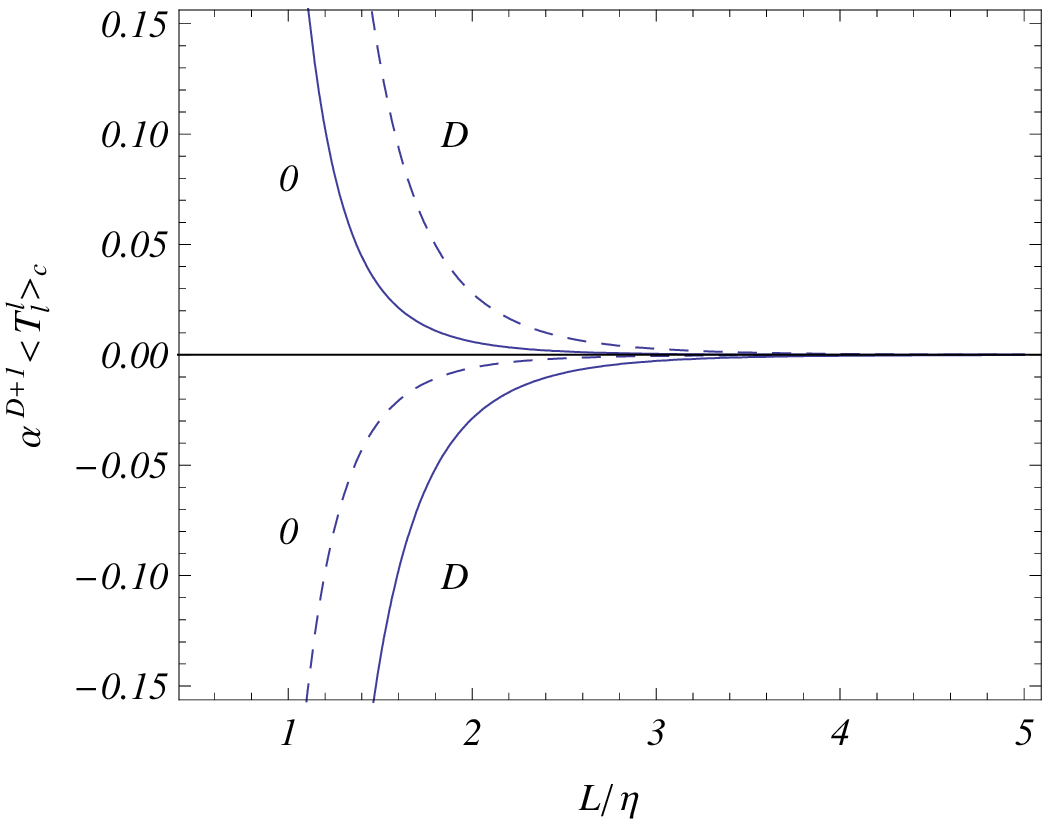,width=7.cm,height=6cm}%
\end{tabular}%
\end{center}
\caption{The topological parts in the VEVs of the energy density
($l=0$) and the vacuum stress along the compactified dimension
($l=D$) in units of the dS curvature scale, $\alpha ^{D+1}\langle
T_{l}^{l}\rangle _{\mathrm{c}}$, as functions of the ratio
$L/\protect\eta $ for the value of the parameter $\protect\alpha
m=1$ and in
the special case of spatial topology $\mathrm{R}^{D-1}\times \mathrm{S}^{1}$%
. Full/dashed curves correspond to spinor fields with
periodicity/antiperiodicity conditions along the compactified
dimension. The left (right) panel corresponds to $D=3$ ($D=4$).}
\label{fig2}
\end{figure}

In figure \ref{fig2} we have plotted the topological parts in the VEVs of
the energy density and the vacuum stress along the compactified dimension
for untwisted (full curves) and twisted (dashed curves) spinor fields as
functions of the ratio $L/\eta $ for the value of the parameter $\alpha m=1$%
. The numbers near the curves indicate the component of the energy-momentum
tensor ($l=0$ for the energy density and $l=D$ for the stress). The left
panel is for $\mathrm{dS}_{4}$ with topology $\mathrm{R}^{2}\times \mathrm{S}%
^{1}$ and the right panel is for $\mathrm{dS}_{5}$ having spatial topology $%
\mathrm{R}^{3}\times \mathrm{S}^{1}$. Note that the ratio $L/\eta =a(\eta
)L/\alpha $ is the comoving length of the compactified dimension in units of
the dS curvature radius. For a massless fermionic field in dS spacetime the
corresponding VEVs are given by formulae (\ref{TlkD3m0}), (\ref{TlkD3m0tw})
and (\ref{TlkR2S1m0}). As we have explained before, in the limit $L/\eta \ll
1$, these expressions are the leading terms in the corresponding asymptotic
expansion for the VEV of the energy-momentum tensor of the massive field.

\begin{figure}[tbph]
\begin{center}
\begin{tabular}{cc}
\epsfig{figure=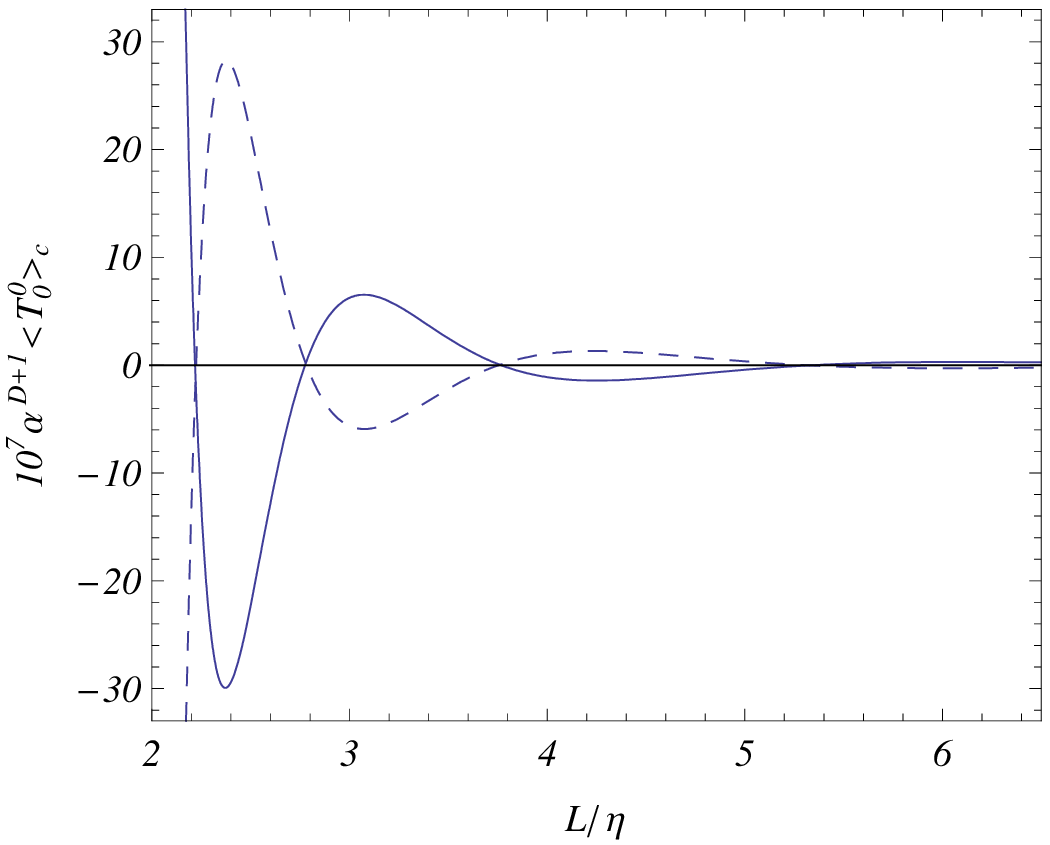,width=7.cm,height=6.cm} & \quad %
\epsfig{figure=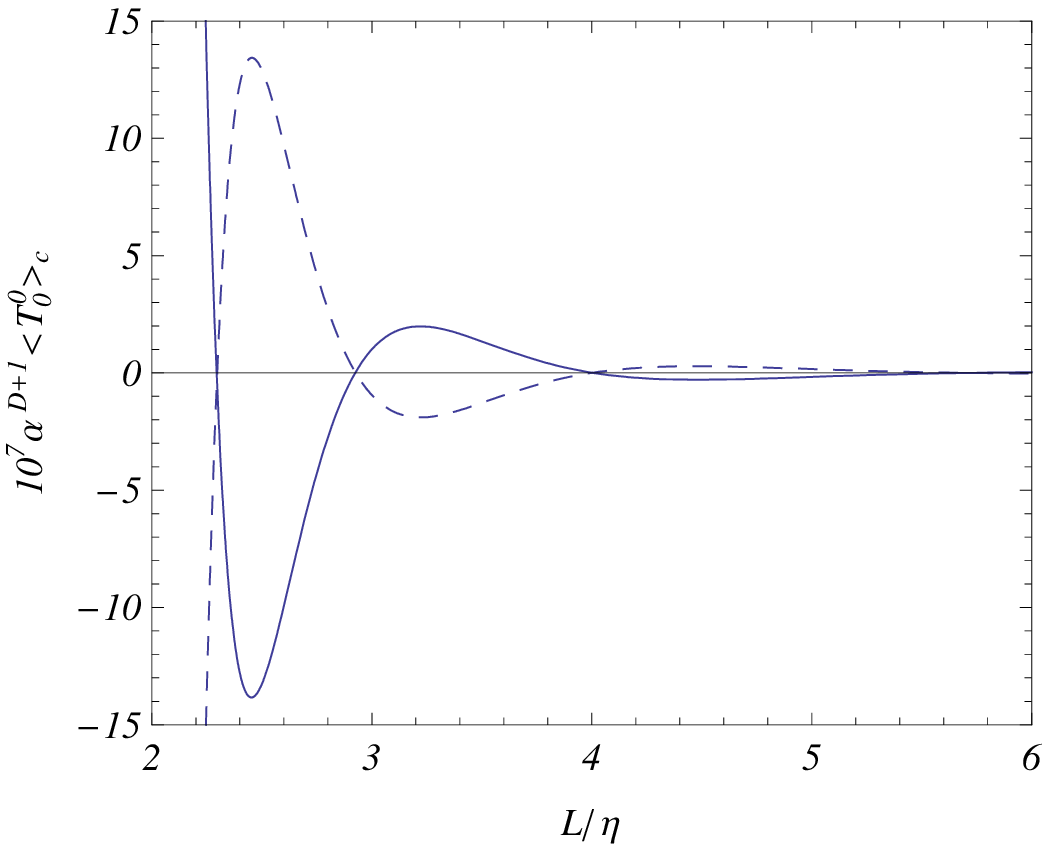,width=7.cm,height=6cm}%
\end{tabular}%
\end{center}
\caption{The topological part in the VEV of the energy density as a function
of the ratio $L/\protect\eta $ for the value of the parameter $\protect%
\alpha m=4$ and in the special case of spatial topology $\mathrm{R}%
^{D-1}\times \mathrm{S}^{1}$. Full/dashed curves correspond to
spinor fields with periodicity/antiperiodicity conditions along
the compactified dimension. The left (right) panel corresponds to
$D=3$ ($D=4$).} \label{fig3}
\end{figure}

From the asymptotic formula (\ref{TllSmall}) it follows that for large
values of the ratio $L/\eta $ the topological part oscillates. As in the
scale of figure \ref{fig2} the oscillations are not well seen, we illustrate
this oscillatory behavior in figure \ref{fig3}, where the topological parts
in the energy density are plotted for untwisted (full curves) and twisted
(dashed curves) fermionic fields $L/\eta $ for $\alpha m=4$. As in figure %
\ref{fig2}, the left panel is for $\mathrm{dS}_{4}$ with topology $\mathrm{R}%
^{2}\times \mathrm{S}^{1}$ and the right panel is for $\mathrm{dS}_{5}$ with
spatial topology $\mathrm{R}^{3}\times \mathrm{S}^{1}$. The first zero with
respect to $L/\eta $ and the distance between neighbor zeros decrease with
increasing values of the parameter $\alpha m$.

\section{Conclusion}

\label{sec:Conc}

We have investigated the Casimir densities for a massive spinor
field in $(D+1)$-dimensional dS spacetime with an arbitrary number
of toroidally compactified spatial dimensions assuming that the
field is prepared in the Bunch-Davies vacuum state. For the
evaluation of the vacuum expectation value of the energy-momentum
tensor the mode-summation procedure is employed. In this approach
we need the corresponding eigenspinors satisfying appropriate
boundary conditions along the
compactified dimensions. These eigenspinors are constructed in section \ref%
{sec:EigFunc} for fields with both periodicity and antiperiodicity
conditions. By using the eigenspinors and applying to the mode-sum the
Abel-Plana formula, the VEV for the spatial topology $\mathrm{R}^{p}\times (%
\mathrm{S}^{1})^{q}$ is presented in the form of the sum of the
corresponding quantity in the topology $\mathrm{R}^{p+1}\times (\mathrm{S}%
^{1})^{q-1}$ and of the part which is induced by the compactness of $(p+1)$%
th dimension. For the field with periodicity conditions the topological part
is given by formula (\ref{TopTll}). The corresponding formula for the field
with antiperiodicity conditions is obtained from that for the field with
periodicity conditions inserting the factor $(-1)^{n}$ in the summation over
$n$ and replacing the definition for $k_{\mathbf{n}_{q-1}}^{2}$ by (\ref%
{knq-1Tw}).

The topological part is finite and the renormalization procedure for the
VEV\ of the energy-momentum tensor is reduced to that for the uncompactified
dS spacetime. This part is time-dependent and breaks the dS symmetry. The
corresponding vacuum stresses along the uncompactified dimensions coincide
with the energy density and, hence, in the uncompactified subspace the
equation of state for the topological part of the energy-momentum tensor is
of the cosmological constant type. For a massless fermionic field the
problem under consideration is conformally related to the corresponding
problem in the Minkowski spacetime with spatial topology $\mathrm{R}%
^{p}\times (\mathrm{S}^{1})^{q}$ and we have the standard relation
$\langle T_{k}^{l}\rangle _{\mathrm{c}}=a^{-(D+1)}(\eta )\langle
T_{k}^{l}\rangle _{\mathrm{c}}^{\mathrm{(M)}}$ between the
topological terms.

For a massive fermionic field, in the limit when the comoving length of a
compactified dimension is much smaller than the dS curvature radius, the
topological part in the VEV of the energy-momentum tensor coincides with the
corresponding quantity for a massless field and is conformally related to
the VEV in toroidally compactified Minkowski spacetime. In particular, the
topological part in the vacuum energy density is positive for an untwisted
fermionic field. This limit corresponds to the early stages of the
cosmological evolution and the topological part dominates over the
uncompactified dS part. At these stages the back-reaction effects of the
topological term are important and these effects can essentially change the
dynamics of the model. In the opposite limit, when the comoving lengths of
the compactified dimensions are large with respect to the dS curvature
radius, in the case of a massive field the asymptotic behavior of the
topological part is damping oscillatory and to the leading order is given by
formula (\ref{TllSmall}). These formula describes the behavior of the
topological part in the late stages of the cosmological expansion. As the
uncompactified dS part is time-independent, we have similar oscillations in
the total VEV as well. Note that this type of oscillatory behavior is absent
for a massless fermionic field.

\section*{Acknowledgments}

The work was supported by the Armenian Ministry of Education and Science
Grant No. 119 and by Conselho Nacional de Desenvolvimento Cient\'{\i}fico e
Tecnol\'{o}gico (CNPq).

\end{document}